\documentclass[conference, a4paper]{IEEEtran}
\usepackage{hyperref}

\IEEEoverridecommandlockouts

	\usepackage[utf8]{inputenc} 
	\usepackage[T1]{fontenc}

\usepackage{cite}

\ifCLASSINFOpdf
  \usepackage[pdftex]{graphicx}
\else
\fi

\usepackage[cmex10]{amsmath}
\usepackage{multirow}
\usepackage{array}
\usepackage[lofdepth,lotdepth]{subfig}
\usepackage{url}
\usepackage{graphicx}
\usepackage{hyperref}

\AtBeginDocument{%
  \renewcommand\tablename{TABLE}
}

\AtBeginDocument{%
  \renewcommand\abstractname{Abstract}
}

\begin{document}

\title{
Integration of Contrastive Predictive Coding and Spiking Neural Networks}

\author{
\IEEEauthorblockN{Emirhan Bilgiç}
\IEEEauthorblockA{
Université Paris-Saclay, France\\
emirhan.bilgic@etu-upsaclay.fr
}
\and
\IEEEauthorblockN{Neslihan Serap Şengör}
\IEEEauthorblockA{
İstanbul Teknik Üniversitesi, Türkiye\\
sengorn@itu.edu.tr
}
\and
\IEEEauthorblockN{Namık Berk Yalabık}
\IEEEauthorblockA{
İstanbul Teknik Üniversitesi, Türkiye\\
yalabik16@itu.edu.tr
}
\and
\IEEEauthorblockN{Yavuz Selim İşler}
\IEEEauthorblockA{
Osmaniye Korkut Ata Üniversitesi, Türkiye\\
yavuzselimisler@osmaniye.edu.tr
}
\and
\IEEEauthorblockN{Aykut Görkem Gelen}
\IEEEauthorblockA{
Erzincan Binali Yıldırım Üniversitesi, Türkiye\\
aykut.gelen@erzincan.edu.tr
}
\and
\IEEEauthorblockN{Rahmi Elibol}
\IEEEauthorblockA{
Hacettepe Üniversitesi, Türkiye\\
rahmielibol@hacettepe.edu.tr
}
}

% make the title area
\maketitle

\begin{abstract}
This study examines the integration of Contrastive Predictive Coding (CPC) with Spiking Neural Networks (SNN). While CPC learns the predictive structure of data to generate meaningful representations, SNN mimics the computational processes of biological neural systems over time. In this study, the goal is to develop a predictive coding model with greater biological plausibility by processing inputs and outputs in a spike-based system. The proposed model was tested on the MNIST dataset and achieved a high classification rate in distinguishing positive sequential samples from non-sequential negative samples. The study demonstrates that CPC can be effectively combined with SNN, showing that an SNN trained for classification tasks can also function as an encoding mechanism. Project codes and detailed results can be accessed on our GitHub page: \href{https://github.com/vnd-ogrenme/ongorusel-kodlama/tree/main/CPC_SNN}{\url{https://github.com/vnd-ogrenme/ongorusel-kodlama/tree/main/CPC_SNN}}
%\boldmath
\end{abstract}
\begin{IEEEkeywords}
predictive coding, contrastive predictive coding, spiking neural networks, self-supervised learning, computer vision
\end{IEEEkeywords}

\IEEEpubid{\makebox[\columnwidth]{\textbf{979-8-3315-6655-5/25/\$31.00 ©2025 IEEE}\hfill}
\hspace{\columnsep}\makebox[\columnwidth]{}}

\IEEEpeerreviewmaketitle

\IEEEpubidadjcol

\section{Introduction}

Predictive coding (PC) is an information processing theory in which higher-level cortical regions of the brain predict sensory inputs from lower levels. According to this theory, learning in the nervous system occurs through hierarchical prediction errors \cite{paper1}. Contrastive predictive coding (CPC) is a self-supervised learning method that makes autoregressive predictions using contrastive loss between positive and negative examples. It learns hidden representations that preserve temporal context \cite{paper2}.

On the other hand, spiking neural networks (SNNs) are an approach that imitates the temporal dynamics of biological neurons and is especially used in event-based information processing models. Information processing occurs through discrete spike-timing rather than continuous activation values \cite{paper3}. Integrating these two methods to better model the processing in the brain will both contribute to biologically inspired modeling and help to develop more energy-efficient systems in practice \cite{paper7}.

The use of CPC and SNNs separately for time series prediction is a subject of research \cite{paper5, paper6}. Studies have been conducted on constructing biologically realistic structures by integrating PC and SNNs \cite{paper7, paper8, paper9}. In addition, SNN structures are trained specifically for encoding tasks and used as autoencoders \cite{paper4, paper10}. However, the encoding capacity of SNNs trained for different tasks has not been tested. While structures combining PC with SNNs have been investigated, to the best of the authors' knowledge, there has been no study specifically integrating CPC with SNNs. In this respect, this work is the first in its field.

In this study, how CPC can be integrated with SNNs is demonstrated, the encoding capacity of SNNs trained on a classification task is examined, and the effects of these integrations are tested on a sequentially constructed MNIST (handwritten digits) dataset. The samples in the MNIST dataset are paired sequentially (and non-sequentially for negative examples) to create training pairs. The designed CPC/SNN integration model is expected to distinguish positive training pairs containing sequentiality from negative training pairs that are not sequential.

\section{Method}
The following steps were carried out during the development of the proposed model:
\begin{itemize}
    \item The encoder structure in CPC, which includes a convolutional neural network, was replaced with SNN to enable encoding with SNNs.
    \item Both SNN trained specifically for classification (SNN-Classifier) and SNN trained specifically for encoding (SNN-Autoencoder) were tested as encoders.
    \item Gated Recurrent Unit (GRU) was used as the autoregressive component in CPC.
    \item The learning mechanism of the neural network was developed by combining Spike-Timing Dependent Plasticity (STDP) rules with the CPC loss.
\end{itemize}

\subsection{Implementation of Encoding with SNN-Classifier}\label{vnd-siniflandirici}
First, instead of an SNN specifically trained for the encoding task, the SNN structure proposed in \cite{paper3}-trained for the 10-class classification problem of the MNIST dataset-was used to encode the MNIST dataset. This structure employs the leaky integrate-and-fire (LIF) cell model and learns via spike-timing-dependent plasticity (STDP).

MNIST images were fed into the model as 784-dimensional input vectors and transmitted to a 400-neuron excitatory layer via a Poisson neuron group. Input data determined the firing rates of neurons, and the model ensured that neurons exceeding specific threshold values generated spikes. By tracking spike activity, a 1x400-dimensional encoding vector was obtained for each image. Finally, spike vectors calculated separately for each digit were saved and combined.

For each pixel \(i\) in the input image, the number of spikes \(s_i\) generated during the time window \(\Delta t\) is modeled using a Poisson distribution, as given in Equation \ref{eq:eq1}:

\begin{equation}\label{eq:eq1}
s_i \sim \text{Poisson}(\lambda_i), \quad \lambda_i = k \cdot I_i \cdot \Delta t
\end{equation}

Here, \(I_i \in [0,1]\): Normalized pixel intensity, \(k\): Scaling factor (\texttt{intensity} parameter), \(\Delta t\): Simulation duration (350 ms), and \(\lambda_i\): Neuron firing rate (Hz).

\subsection*{Minimum Spike Condition}
If the total number of spikes generated \(S_{\text{total}} = \sum_{i=1}^{784} s_i\) falls below a certain threshold (\(S_{\text{total}} < S_{\text{min}}\)), the scaling factor \(k\) is adaptively increased:

\begin{equation} \label{eq:eq2}
    k \leftarrow k + \Delta k \quad \text{if} \quad S_{\text{total}} < S_{\text{min}}
\end{equation}

The following values were used during simulation:  
\(\Delta k = 1\), \(S_{\text{min}} = 5\) spikes, and \(\Delta t = 0.35\) seconds.

\vskip 0.2in

\begin{figure}[h]
    \centering
    \begin{tabular}{cc}
        \includegraphics[scale=0.18]{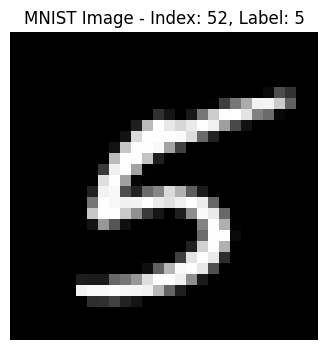} & \includegraphics[scale=0.18]{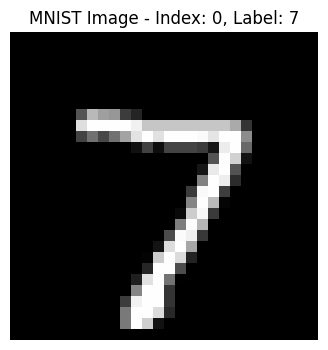} \\
        \includegraphics[scale=0.18]{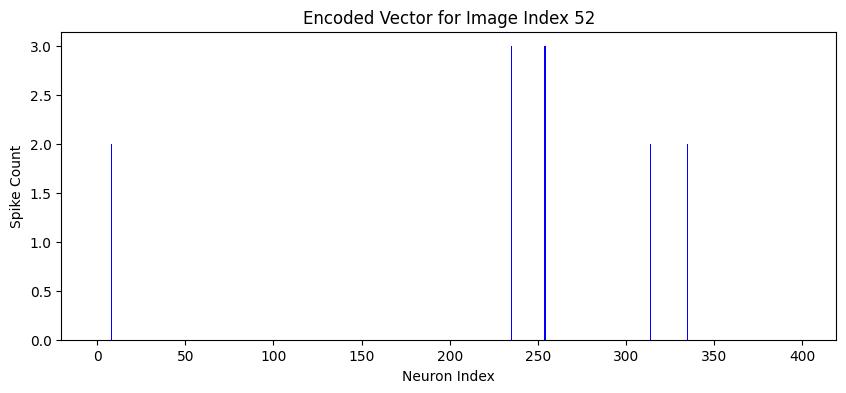} & \includegraphics[scale=0.18]{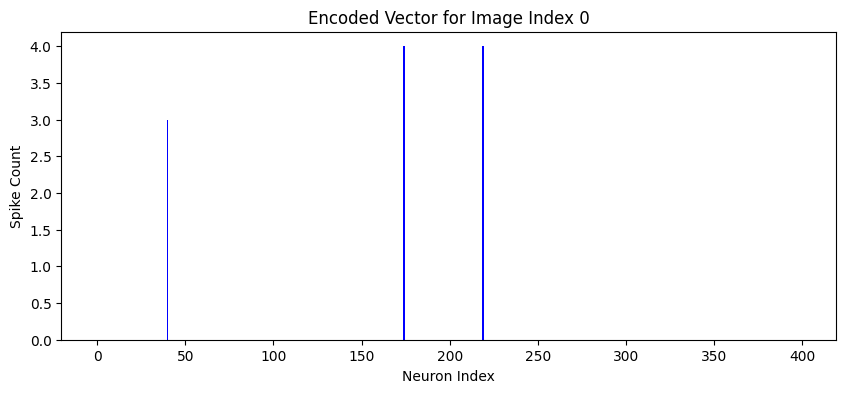} \\
        (a) 5 and SNN-Encoded 5 & (b) 7 and SNN-Encoded 7 \\
        [0.4cm]
         \includegraphics[scale=0.18]{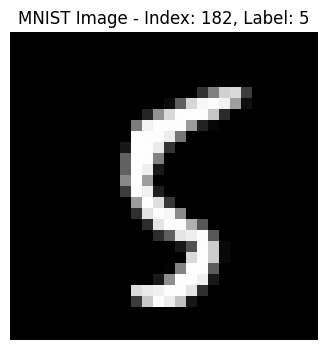}&
         \includegraphics[scale=0.18]{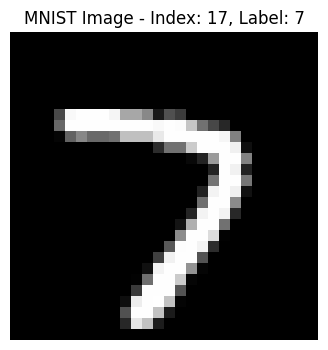} \\
         \includegraphics[scale=0.18]{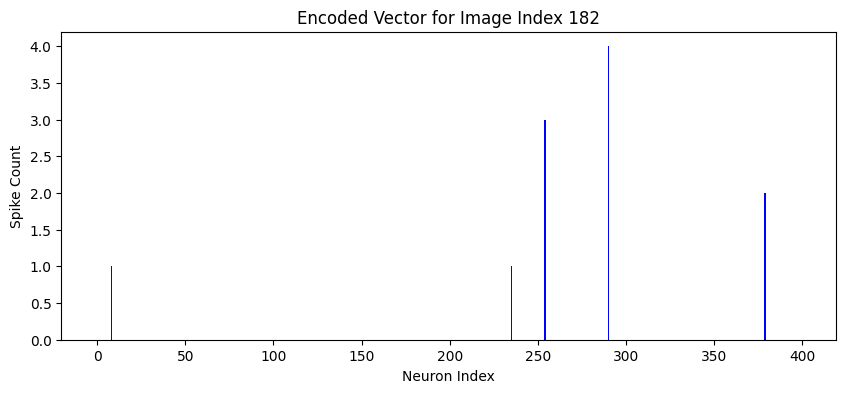}&
         \includegraphics[scale=0.18]{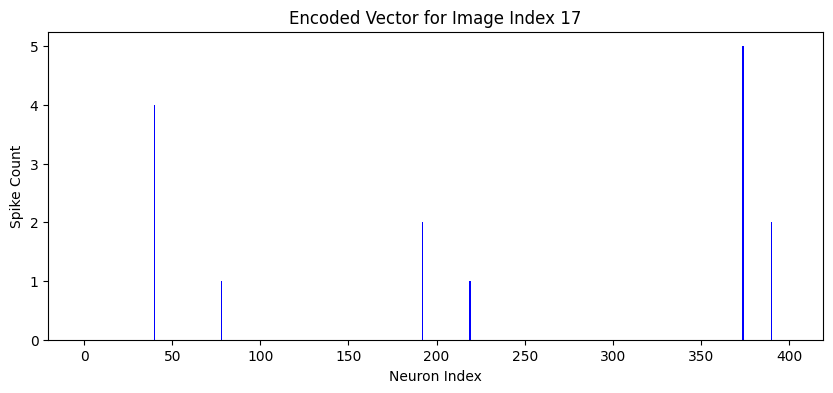} \\
        (c) 5 and SNN-Encoded 5 & (d) 7 and SNN-Encoded 7 \\
    \end{tabular}
    \caption{Visualization of 400x1-dimensional vectors generated by the SNN, representing spike counts per neuron for the digits 5 and 7. These vectors were created for encoding purposes.}
    \label{kodlama}
\end{figure}

The vectors visualizing spike counts for different digits exhibit distinct characteristics. Within-class computations (e.g., only for the digit 5) produce patterns with visually noticeable similarities. Figure~\ref{kodlama} shows only two digits and their SNN-encoded counterparts; a detailed table is available on our GitHub page.

\subsection{Implementation of Encoding with SNN-Autoencoder}\label{vnd-otokodlayici}
As mentioned in Section \ref{vnd-siniflandirici}, after encoding with an SNN trained specifically for classification and based on the leaky integrate-and-fire model \cite{paper3}, an autoencoder-also based on the LIF model but trained explicitly for the encoding task-was developed for comparison.

\subsection{Similarity Calculation (Dot Product and Average)}

The task of the SNN autoencoder here is not classification but minimizing the reconstruction error between the original (uncoded) data and the decoded data after encoding. Since this structure is trained specifically for encoding, it is expected to exhibit higher encoding capability and consequently achieve higher success when integrated with CPC. However, this encoding relies on convolutional neural networks (CNNs), making it less biologically realistic compared to the encoding in Section \ref{vnd-siniflandirici}. Although the convolutional layers in the SNN-Autoencoder mimic the locally connected structure of biological neural networks, they diverge from true neuronal dynamics (e.g., spike-timing-based communication).

The encoding process with the autoencoder is performed over 25 time steps. At each step, the membrane potentials of neurons are updated, and the membrane potential of the final time step is used as the latent vector. For integration with CPC, only the encoder part's weights were frozen and utilized.

The dynamics of the leaky integrate-and-fire (LIF) model are described by Equations \ref{eq:eq_lif} and \ref{eq:eq_lif2}:

\begin{equation}\label{eq:eq_lif}
    V[t] = \underbrace{\beta V[t-1]}_{\text{Leakage}} + \underbrace{(1-\beta)(W_k \ast X)}_{\text{Input Contribution}}
\end{equation}

\begin{equation}\label{eq:eq_lif2}
    S[t] = 
    \begin{cases} 
        1, & V[t] \geq V_{thresh} \\
        0, & \text{otherwise}
    \end{cases}
\end{equation}

Here, \(W_k\): Weight matrix of the \(k\)-th layer,  
\(\beta\): Membrane potential leakage factor,  
\(X\): Input activations.

\subsection{Similarity Calculation (Dot Product and Average)}

The similarity between the true encoded vectors (\(y\)) and the predicted encoded vectors (\(p\))-generated by the CPC network using a combination of GRU and dense neural networks as shown in Figure~\ref{sekil_cpc_vnd}-is calculated via the dot product method. During training, the CPC network aims to increase the dot product between predicted and encoded vectors for positive examples while decreasing it for negative examples. During testing, whether a new data sequence is positive or negative is determined by calculating the normalized dot product between predicted and encoded vectors: a high score (>0.5) indicates a positive example, while a low score (<0.5) indicates a negative example.

The similarity at each time step \(t\) is expressed by Equation \ref{eq:eq3}:
\begin{equation}\label{eq:eq3}
    s_t = \sum_{i=1}^d p_t[i] \cdot y_t[i]
\end{equation}

Where \(p_t\): Predicted encoding vector,  
\(y_t\): True encoding vector,  
\(d\): Dimension of the encoding vector,  
\(s_t\): Similarity score at time \(t\).

These scores are then averaged:
\begin{equation}\label{eq:eq4}
    \text{Mean Score} = \frac{1}{T} \sum_{t=1}^T s_t
\end{equation}

Where \(t\): time step.

This represents the overall similarity between predicted and true vectors across all time steps. The mean similarity score is also called the raw score or logit. The logit value is normalized to the \([0,1]\) range using the sigmoid activation function to measure prediction accuracy.

\subsection{Binary Cross-Entropy Loss Function}

The model's output (logits passed through sigmoid activation) is compared with the true labels to calculate the binary cross-entropy (BCE) loss. The loss formula for positive (\(1\)) and negative (\(0\)) examples is given by Equation \ref{eq:eq6}:
\begin{equation}\label{eq:eq6}
    \text{BCE} = - \frac{1}{N} \sum_{i=1}^N \left[ y_i \cdot \log(\hat{y}_i) + (1 - y_i) \cdot \log(1 - \hat{y}_i) \right]
\end{equation}

Here, \(N\): Number of samples, \(y_i\): True label (\(0\) or \(1\)), \(\hat{y}_i\): Predicted probability (\(\hat{y}_i = \text{sigmoid}(\text{logit})\)).

\begin{figure}[h]
	\centering
	\includegraphics[scale=0.45]{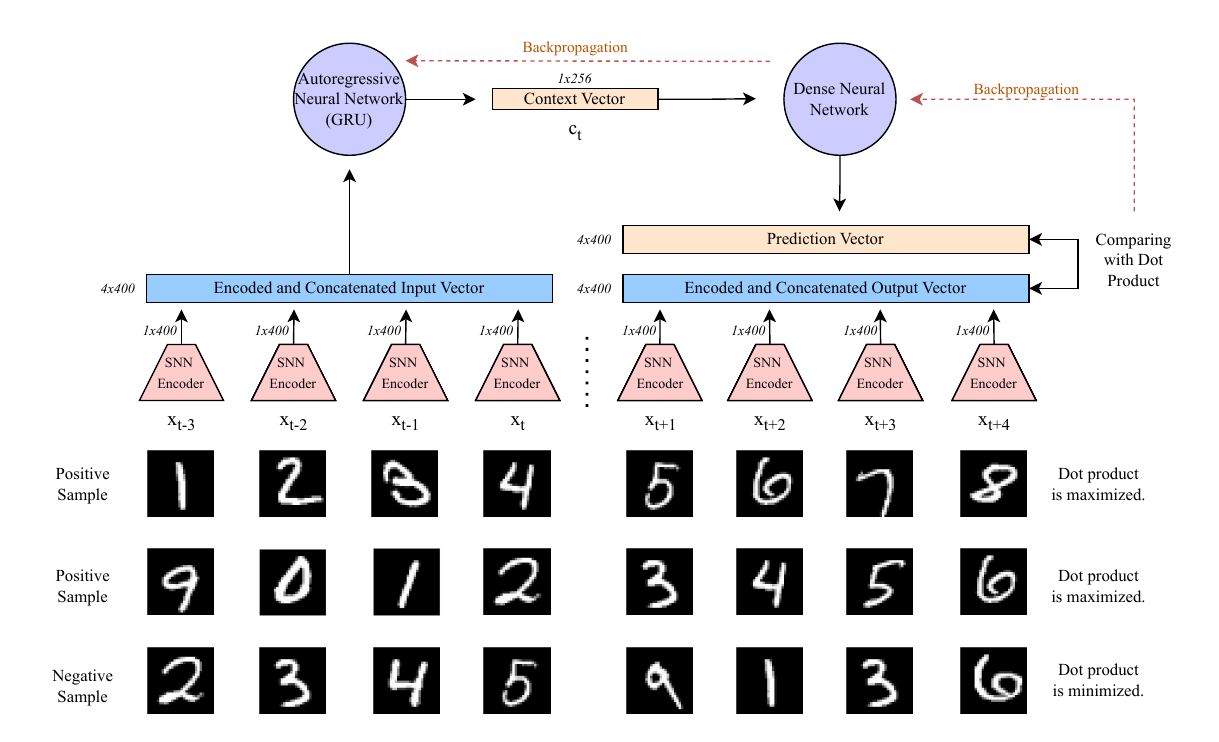}
	\caption{Structure of the proposed CPC and SNN integration model. The first example ([1, 2, 3, 4] -> [5, 6, 7, 8]) is a positive sample as it contains two sequential number sequences, while the second example is another positive sample. The last example is a negative sample.}
	\label{sekil_cpc_vnd}
\end{figure}

This loss is minimized via backpropagation to optimize the prediction performance of the dense neural network in the model, as shown in Figure~\ref{sekil_cpc_vnd}.

\section{Training Details and Model Performance}

Two subsets of the MNIST dataset were created for training:
\begin{enumerate}
    \item 250 samples per class (10 classes), totaling 2500 samples.
    \item 500 samples per class (10 classes), totaling 5000 samples.
\end{enumerate}

Encoding was performed using both:
\begin{enumerate}
    \item The SNN-Classifier described in Section~\ref{vnd-siniflandirici} (trained for classification).
    \item The SNN-Autoencoder described in Section~\ref{vnd-otokodlayici} (trained explicitly for encoding).
\end{enumerate}

The training dataset used batches of 32 positive and 32 negative examples, while the validation dataset used batches of 10 positive and 10 negative examples. Training employed the Adam optimizer (\(\eta=10^{-4}\)) and BCE loss. Training ran for a maximum of 100 epochs, with early stopping if validation accuracy did not improve for 10 epochs. If validation loss did not improve for three epochs, the learning rate was halved.

Each training run was repeated three times, with results reported as means ± standard deviations. For example, if the mean accuracy at an epoch was 0.85 with a standard deviation of 0.03, the graph shaded the region between 0.82 and 0.88. Shading was truncated if some trials ended earlier than others.

\begin{figure}[h]
	\centering
	\includegraphics[scale=0.3]{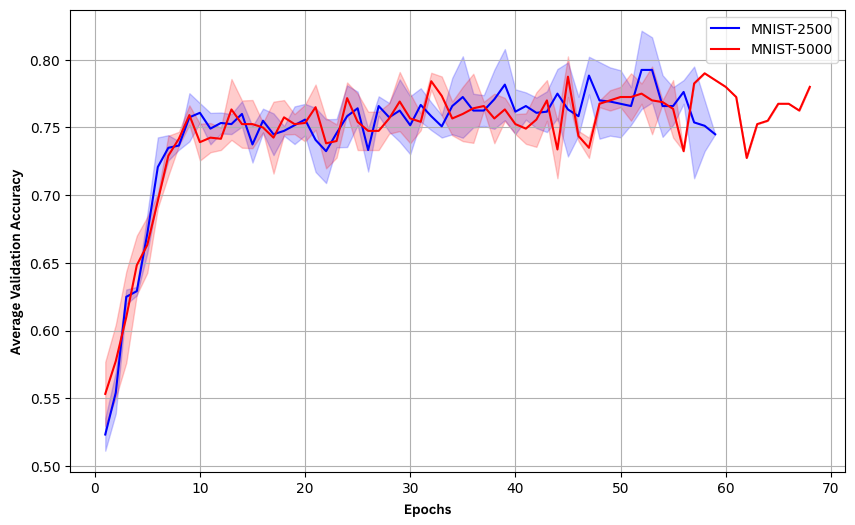}
	\caption{Validation accuracy graph of the CPC network using SNN-Classifier for encoding.}
	\label{vnd_siniflandirici_basarisi}
\end{figure}

\begin{figure}[h]
	\centering
	\includegraphics[scale=0.26]{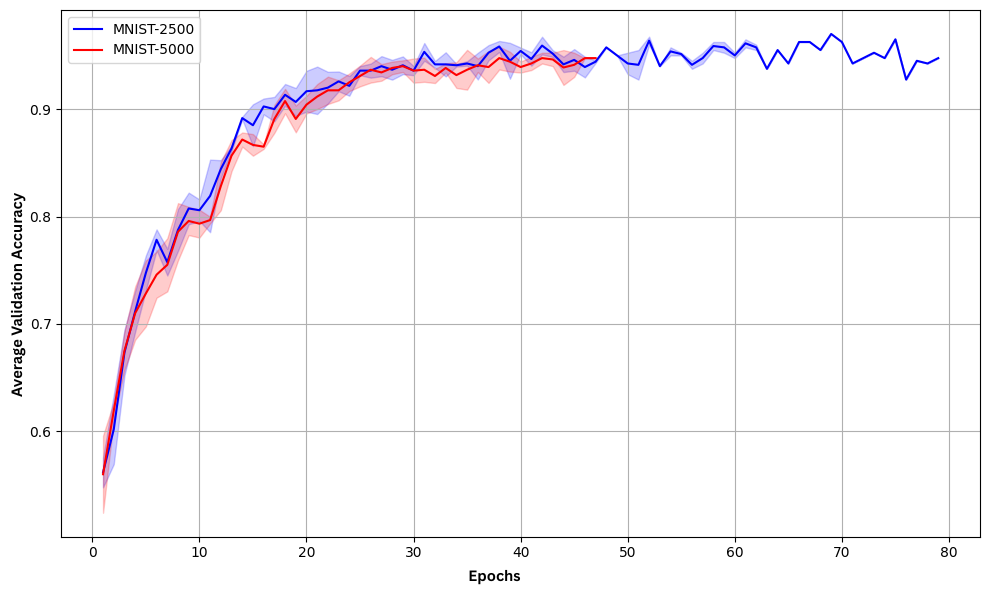}
	\caption{Validation accuracy graph of the CPC network using SNN-Autoencoder for encoding.}
	\label{vnd_otokodlayici_basarisi}
\end{figure}

To test the model's validity, MNIST images were randomly encoded (instead of using SNNs) to evaluate the classification accuracy in the CPC network.

\begin{figure}[h]
	\centering
	\includegraphics[scale=0.3]{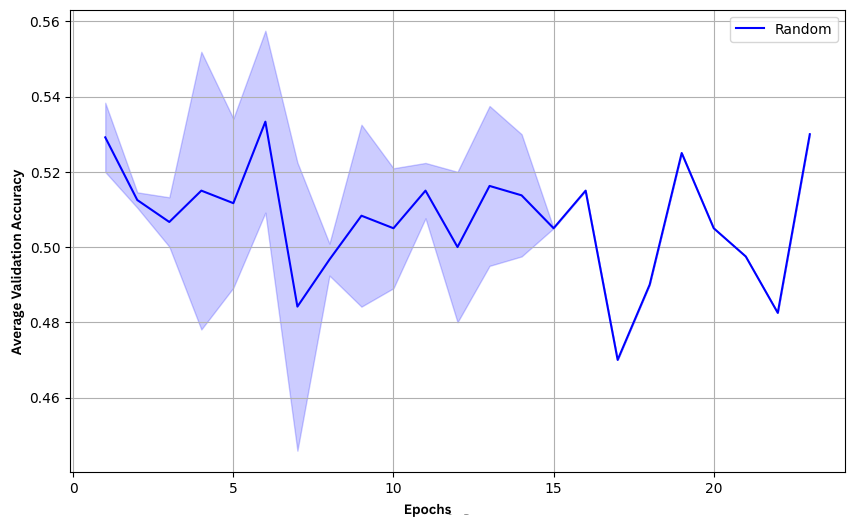}
	\caption{Validation accuracy graph of the model with random encoding.}
	\label{basarim_random}
\end{figure}

\begin{table}[h]
  \centering
  \caption{\textsc{Training Results of the CPC Network}}
  \label{tablo1}
  \begin{tabular}{|c|c|c|c|}
    \hline
    Dataset & Encoding Method & Max Validation Accuracy & Epoch \\
    \hline
    MNIST-2500  & SNN-Autoencoder       & 0.9683 & 62.67 \\
    MNIST-5000  & SNN-Autoencoder       & 0.9583 & 44.33 \\
    MNIST-2500  & SNN-Classifier    & 0.8033 & 57.33 \\
    MNIST-5000  & SNN-Classifier    & 0.7992 & 55.00 \\
    MNIST-2500  & Random  & 0.5558 & 16.00 \\
    \hline
  \end{tabular}
\end{table}

Table \ref{tablo1} presents results averaged over three independent training runs.

As shown in Table \ref{tablo1}, the CPC network using the SNN-Autoencoder (trained specifically for encoding) achieved the highest validation accuracy for positive/negative sample classification. Increasing the dataset size did not significantly improve validation accuracy but resulted in earlier stopping (i.e., training concluded sooner when validation accuracy plateaued for 10 epochs).

While the SNN-Autoencoder outperformed the SNN-Classifier, the latter - as explained in Section \ref{vnd-otokodlayici} - remains more biologically plausible. Nevertheless, the SNN-Classifier demonstrated reasonable success, proving that biologically realistic SNNs trained for classification tasks can also serve as encoders and integrate effectively with CPC.

The CPC network with random encoding performed near chance level ($\sim$55\%), confirming the validity of the proposed model.

\section{Conclusions and Recommendations}

This new approach has presented a model that is both more aligned with biological principles and has high learning capacity. The proposed model has shown that both CPC and SNN structures can be used together, as well as demonstrating that VND trained specifically for a classification task can be used in an encoding task.  

In future work, the model can be tested on other time-dependent datasets (e.g., Human Action Recognition). Furthermore, instead of CPC, the structure from \cite{paper1}, which is biologically more realistic, can be used, and it can be tested on neuromorphic hardware.

\section*{Acknowledgments}
This study was supported by TÜBİTAK project number 23E674.


\begin{thebibliography}{1}

\bibitem{paper1}
Rao, R. P., \& Ballard, D. H. (1999). Predictive coding in the visual cortex: a functional interpretation of some extra-classical receptive-field effects. Nature neuroscience, 2(1), 79-87.

\bibitem{paper2}
Oord, A. V. D., Li, Y., \& Vinyals, O. (2018). Representation learning with contrastive predictive coding. arXiv preprint arXiv:1807.03748.

\bibitem{paper3}
Diehl, P. U., \& Cook, M. (2015). Unsupervised learning of digit recognition using spike-timing-dependent plasticity. Frontiers in computational neuroscience, 9, 99.

\bibitem{paper4}
Comşa, I. M., Versari, L., Fischbacher, T., \& Alakuijala, J. (2021). Spiking autoencoders with temporal coding. Frontiers in neuroscience, 15, 712667.

\bibitem{paper5}
Deldari, S., Smith, D. V., Xue, H., \& Salim, F. D. (2021, April). Time series change point detection with self-supervised contrastive predictive coding. In Proceedings of the web conference 2021 (pp. 3124-3135).

\bibitem{paper6}
Lucas, S., \& Portillo, E. (2024). Methodology based on spiking neural networks for univariate time-series forecasting. Neural Networks, 173, 106171.

\bibitem{paper7}
N'dri, A. W., Gebhardt, W., Teulière, C., Zeldenrust, F., Rao, R. P., Triesch, J., \& Ororbia, A. (2024). Predictive Coding with Spiking Neural Networks: a Survey. arXiv preprint arXiv:2409.05386.

\bibitem{paper8}
Lee, K., Dora, S., Mejias, J. F., Bohte, S. M., \& Pennartz, C. M. (2024). Predictive coding with spiking neurons and feedforward gist signaling. Frontiers in Computational Neuroscience, 18, 1338280.

\bibitem{paper9}
Boerlin, M., Machens, C. K., \& Denève, S. (2013). Predictive coding of dynamical variables in balanced spiking networks. PLoS computational biology, 9(11), e1003258.

\bibitem{paper10}
Auge, D., Hille, J., Mueller, E., \& Knoll, A. (2021). A survey of encoding techniques for signal processing in spiking neural networks. Neural Processing Letters, 53(6), 4693-4710.

\end{thebibliography}
\end{document}